# After 60 years, here comes the Sun

Anatael Cabrera[1,2,*], Mark Chen[3], Stefano Dusini[4], Jeff Hartnell[5],
Jorge Martín Camalich[6,7], Diana Navas[8], Hiroshi Nunokawa[9],
Aldo Serenelli[10], Javier Trujillo Bueno[6,7,11], Frédéric Yermia[12]

[1] *IJCLab, CNRS, Université Paris-Saclay, France*
[2] *LNCA, CNRS-EDF, Chooz Nuclear Reactor, France*
[3] *Department of Physics, Engineering Physics & Astronomy, Queen's University, Kingston, Canada*
[4] *INFN, Sezione di Padova, Italy*
[5] *Department of Physics and Astronomy, University of Sussex, Brighton, United Kingdom*
[6] *Instituto de Astrofísica de Canarias, Tenerife, Spain*
[7] *Departamento de Astrofísica, Facultad de Física, Universidad de La Laguna, Tenerife, Spain*
[8] *Centro de Investigaciones Energéticas, Medioambientales y Tecnológicas, Spain*
[9] *Departamento de Física, Pontifícia Universidade Católica do Rio de Janeiro, Brazil*
[10] *Institute of Space Sciences (ICE, CSIC), Spain*
[11] *Consejo Superior de Investigaciones Científicas, Spain*
[12] *SUBATECH, CNRS, IMT Atlantique, Université de Nantes, France*

For sixty years, the Sun served neutrino physics more than the reverse: its neutrinos established oscillations rather than probing the star itself. That is now changing. Reactor antineutrinos have pinned down the oscillation parameters governing solar-neutrino flavour conversion so precisely that the propagation-related uncertainty is becoming negligible — closing the first of two abysses. Solar neutrinos can, at last, be turned on the Sun. Yet a second abyss remains: today's solar-neutrino detection falls an order of magnitude short of the precision the Standard Solar Model already predicts for the dominant pp flux (>90%) — a gap no reactor experiment can close. The intrinsic $^{14}C$ background walls off the pp region; only new detection techniques can breach it. Bridging this second abyss demands a revolution, not an increment. Combined with helioseismology's complementary reading of the Sun's mechanical structure, precision neutrino measurements of the nuclear source terms would enable a multi-messenger heliotomography of the solar interior — reading our star from both sides at once. A new age of solar discoveries may follow, on one condition: that the field builds the instruments equal to the task.

* anatael@in2p3.fr

## The pioneers' ambition: on hold for sixty years — so far

For about six decades, the Sun has served neutrino physics more than the reverse — less than the pioneers dreamt. What began in the 1960s as *Davis*'s audacious attempt to observe the solar interior through *Bahcall*'s predicted neutrino flux became one of the most consequential anomalies in modern physics: the "solar neutrino problem" **[1]**. Its resolution — the quantum oscillation of neutrino flavour, established over several decades and culminating in Super-Kamiokande **[2]**, SNO **[3]** and KamLAND **[4]** experiments — earned the field two Nobel Prizes **[5]** and delivered the concrete laboratory evidence of physics beyond the Standard Model of Particle Physics (SMPP): neutrinos are massive. The riddle of their mass, however, remains unsolved, though the combination of many experiments is now beginning to settle the so-called neutrino mass ordering — an important ingredient for interpreting future absolute-mass measurements **[6]**.

While nothing known at the time required neutrinos to be massless, this assumption was later adopted as a building block in the SMPP Lagrangian. Hence, oscillations already demand an extension of that minimal theory **[7]**. Between exactly massless and merely light lies a gap in interpretation, perhaps hinting at a new symmetry. That minute difference may shape the Universe from its earliest epoch to today as massive neutrinos leave measurable imprints on cosmological structure formation, and cosmology now provides powerful — though model-dependent — constraints on their possible role, including the sum of all neutrino masses.

The establishment of oscillations has, to our knowledge today, completed our picture of neutrinos' propagation through space and time, as relativistic nomadic particles. But there was a price, largely unspoken: the fixation on settling that propagation question monopolised the narrative for a long time, deferring what the pioneers first imagined — to see uniquely into the deepest interior of matter, including inside our own star, and turn that view into a new astrophysics. Indeed, if we are to understand our existence, the Sun is one of the brightest places to look first.

So, for decades, the Sun was treated mainly as a source — the most copious natural neutrino source reaching Earth. The propagation of those neutrinos to us is primarily governed by two oscillation parameters, $\theta_{12}$ and $\Delta m^2_{21}$, known historically as the "solar parameters". The propagation probability is also strongly shaped by matter effects within the Sun and, to a lesser extent, within the Earth, and is slightly dependent on $\theta_{13}$[1].

Our Sun is, unsurprisingly, the most studied star in history, with its knowledge captured not in a single object but in a calibrated framework: the Standard Solar Model (SSM), a family of one-dimensional evolutionary models tuned to the Sun's luminosity, radius, age and surface composition **[8,9]**. Spherically symmetric by construction, the SSM has limitations in known physically present processes, such as rotation, angular momentum transport, rotationally induced chemical mixing, or magnetic field effects **[10]**. Any discrepancy with future high-precision data could also reveal the missing transport physics, opening an additional avenue for discovery.

Neutrinos have already made important contributions to it. Indeed, neutrino observations have confirmed that the pp chain is the primary driver and that the CNO cycle (Carbon-Nitrogen-Oxygen) operates in the Sun, as shown by Borexino **[11]**. Moreover, the SNO's neutral-current measurement showed that the total active $^8$B flux matched the SSM prediction, confirming its accuracy for decades and that the solar-neutrino deficit was due to flavour conversion — signature evidence of neutrino oscillations during propagation. Today, the solar core temperature is most sharply constrained by neutrinos: the steep temperature dependence of the $^8$B flux ($\Phi \propto T_c^{24}$) yields the deep-core temperature to ~0.7% **[12]**, a precision limited by the nuclear inputs in the relation between flux and temperature **[13]**.

[1] All other elements of the standard neutrino oscillation model, still under exploration, do not affect the propagation of the solar neutrinos from the core of the Sun to Earth. The main elements are: the existence of leptonic CP-violation and the resolution of both the so-called mass ordering and the octant-$\theta_{23}$.

What neutrinos have not yet done, though, is reach the experimental precision needed to refine the SSM across the full energy spectrum. What started with the combined solar-KamLAND analysis **[14]** has transformed into a sustained phenomenological programme that has developed the statistical methodology **[15,16,17,18]** to turn neutrino measurements into SSM constraints, i.e., a data-driven reconstruction of solar properties. That effort has shown what the data can already say **[17]** and built the inferential tools any future experiment will need **[18]**. What remains is the experimental leap: the frontier ahead requires measurement precision that approaches or surpasses the SSM's own predictions. Neutrinos are the only messenger that reports, essentially instantaneously — after the ~8-minute flight to Earth — on what is happening right now in the core. This is in stark contrast to the emergence of energy carried outward by photons, delayed for tens of thousands of years after diffusion **[19,20]**.

Their still-underexploited viewpoint is fully complementary to the heavily developed astronomical paradigm, which reads the Sun from the outside inwards: photons mapping the surface, helioseismology propagating that information inward through the Sun's global acoustic oscillations **[20]**. The two messengers, neutrinos and sound, are complementary, not because one reaches the interior and the other does not — global acoustic modes penetrate deeply — but because they constrain different phenomena: while helioseismology measures the Sun's mechanical and thermodynamic structure across most of its radius, neutrinos measure the nuclear fusion source terms — the reaction rates — in the core itself **[8]**. When those two visions meet at comparable precision, many surprises — discoveries — are expected.

## The power of neutrinos on Earth: sixty days with reactors

That wait is closing. First released in November 2025 and published in June 2026 **[21]**, the JUNO experiment (based in China) used less than 60 days of reactor-antineutrino data to determine the two "solar parameters" more precisely than the previous global combination — six decades of solar-neutrino data included. The published values ($\sin^2\theta_{12} \approx 0.309 \pm 0.009$ and $\Delta m^2_{21} \approx (7.50 \pm 0.12)\times10^{-5}$ eV$^2$) improve the prior knowledge by a factor of roughly 1.6×. The reactor route's pioneer, the KamLAND experiment (based in Japan), is likewise surpassed. Sixty days on Earth, once tuned to ultimate precision, have overtaken sixty years of looking at the sky.

A JUNO update presented at Neutrino-2026 **[22]** indicates further improvement toward the expected ~1% level, as the experiment was designed to reach few-permille precision in both parameters with full exposure ultimately **[23]**, if systematics go as planned. What matters is not the headline but its implication: the leading oscillation-parameter contribution to the solar-neutrino uncertainty budget is poised to become negligible. For the first time, the messenger's terrestrial journey is well enough understood that the particle-physics terms no longer limit what solar neutrinos can tell us about their source. Unless discoveries prove otherwise, this closes the first abyss — propagation — bridged on Earth by reactors and antineutrinos.

## From reactors to the most exotic laboratory nearest to Earth

That the parameters governing solar-neutrino propagation are pinned down with antineutrinos from power reactors is no accident: the same mixing angle and mass splitting govern both channels. Ultimate precision belongs to the best-controlled setups, and reactors are exquisite: intense, well-localised antineutrino sources at accurately surveyed baselines. The Sun is the opposite — a magnificent, wild and uncontrolled source, arguably the most exotic physical environment within reach of Earth, whose interior can be read only from within through its neutrinos **[20]** — a viewpoint no telescope so far has reached in full capacity.

Much of JUNO's spectral-extraction precision rests on an exceptionally well-known baseline configuration and unprecedented energy control — linearity at the few-permille level — and on the internal redundancy system, based on the first dual-calorimetry photodetection architecture, which cross-calibrates the energy scale in situ **[24]**; a design our teams pioneered.

## The renaissance of solar neutrinos

Since the triumphs of SNO and KamLAND in the early 2000s, the frontier moved on. Solar neutrinos, once the workhorse of the field, quietly ceased to be its leading target; no new flagship experiment has been designed or built mainly for them for a long time, and several promising projects were discontinued. Indeed, as budgets tighten, they are among the first to be made casualties.

The reason is brutally simple: detecting them remains one of humankind's hardest measurements — and achievements. They are very low in energy, hence even less likely to interact. Unlike reactor antineutrino detection, solar neutrino detection has long sought a comparably clean "prompt-delayed" signature with an abundant target for high-rate, low-threshold operation. Indeed, since the neutrino discovery, reactor antineutrinos have given physicists a golden fingerprint: *inverse beta decay* (IBD) on free protons, whose *double-flash* signature (prompt-then-delayed) dramatically reduces backgrounds, as *Reines* and *Cowan* proved in 1956 **[25]** (Nobel Prize 1995). Existing high-statistics low-energy solar-neutrino measurements generally lack an event-by-event double-flash coincidence tag, so each event is a lonely spark in a sea of radioactivity — γ-rays and β particles — and cosmic-muon-induced nuisance. Overcoming that loneliness by finding novel, tagged, and detectable interactions remains a central challenge of the field.

## Beyond the boundaries of today's neutrino detection technology

Whenever detection limits our ability to probe Nature, the bottleneck is usually the instrument — and here lies the second abyss. Even granting JUNO's terrestrial precision, solar neutrino detection today falls short of the SSM's predictions by roughly an order of magnitude. The gap is not uniform (**Figure 1**): for the temperature-sensitive $^{8}$B flux, the experimental precision already exceeds the model's, but that only sees ~0.01% of the Sun's total flux. Instead, the dominant lowest-energy pp flux (~91.5% of the total) is predicted by the SSM to ~0.6%, because the model is calibrated to the solar luminosity. In comparison, the experimental precision is on the order of ~10% — a gap of nearly 20×. Here, no reactor experiment, however precise, can close this second abyss: it is a limitation of the direct solar neutrino detection and instrumentation, not of neutrino physics knowledge. The main culprit limiting precision is the intrinsic $^{14}$C background — the same one used for archaeological dating — which has effectively formed an impenetrable wall in the pp region of organic liquid scintillator detectors to date. Indeed, only new techniques may resolve the $^{14}$C challenge.

The way past that wall is not shielding (or passive background rejection), as this is part of the detector composition, but enhancing the information per event: reading, or even imaging, the shape of an interaction rather than only its energy. That is the *raison d'être* of unsegmented high-granularity imaging in scintillator, as pioneered (invented in 2012) by the LiquidO opaque-medium detection technique **[26]**, which trades transparency for localisation and per-event topology — an identity card per event. Hence, LiquidO's advantage is active topological rejection, which promises to bridge the second abyss – a bridge still under exploration. The potential is such that even some passive shielding, including overburden, may be reduced. However, even LiquidO, alone, cannot resolve the single-flash of a solar neutrino, which radiogenic topologies — the overwhelming β background — could still mimic. But it changes what may become possible next.

Some of the most powerful "old" ideas for solar detection outran the instruments of their day. In 1976, *Raghavan* proposed indium (dominated by $^{115}$In isotope) as the interaction target, leading to a unique route to low-threshold, flavour-specific, event-by-event spectroscopy with direct ability to measure the pp solar neutrino via a double-flash coincidence tag **[27]** — this is the matter-sector analogue of *Reines*'s antimatter-sector signature. Indeed, indium is one of the very few known paths to real-time spectroscopy of the lowest-energy solar neutrinos across the full spectrum. Most importantly, the coincidence is defined to render the $^{14}$C wall experimentally surmountable for the first time, by construction.

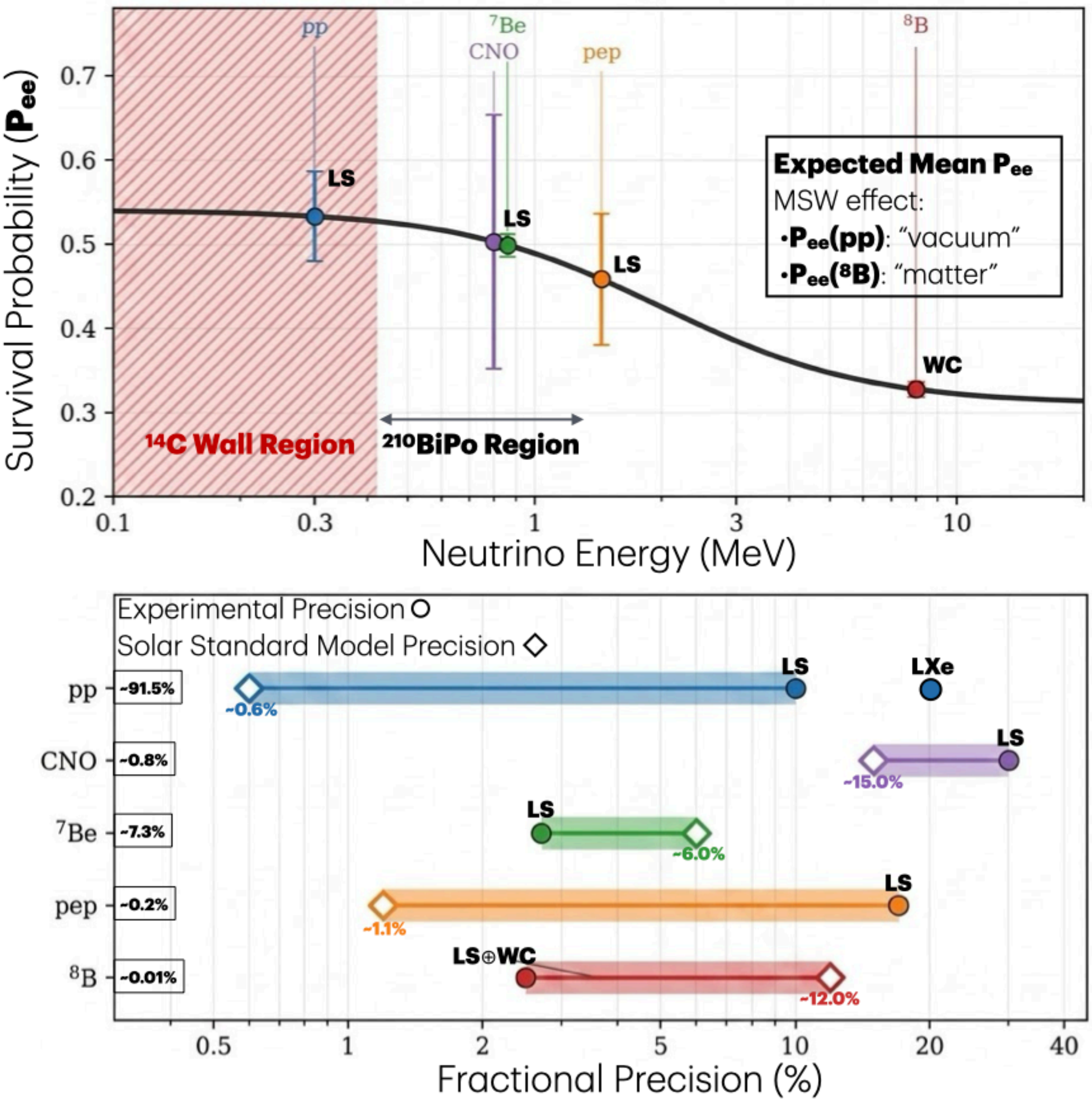

**Figure 1 | Solar-neutrino precision today, and the $^{14}$C wall on pp.** Today's knowledge on the electron-neutrino survival probability ($P_{ee}$) as a function of the energy **(top)** is illustrated with the principal solar reaction, the leading pp-chain (subcomponents: pp, $^7$Be, pep, $^8$B) and the sub-dominant CNO-cycle (barely ~1% of the total), including their current knowledge and experimental precision (error-bars) as measured by neutrinos. The $P_{ee}$ transits from "vacuum" (low energies: pp) to "matter" (high energies: $^8$B), hypothetically dominated by the MSW effect (*Mikheyev-Smirnov-Wolfenstein*), whose transition remains poorly probed by experiments, so that other effects may occur. The leading probing technology is *liquid scintillator* (**LS**), but the high-energy range can also be accessed with *water Cherenkov* (**WC**) detectors. The so-far dominant $^{14}$C (shaded region), affecting mainly organic scintillator detectors, limits the ultimate precision of the pp-component in transparent detectors. The $^{14}$C limitation may be, however, resolved using other techniques: i) detectors not made of carbon and ii) a carbon detector using coincidences, such as indium. While a priori solvable, today the CNO is limited by the $^{210}$Bi-$^{210}$Po background contamination. Still low precision, the latest measurement of pp by *liquid xenon* dark matter experiments (**LXe**) may also be impacted by $^{14}$C contamination. The fractional precision per component is illustrated **(bottom)**: the SSM prediction (diamonds and numerical value provided) versus the current best experiment (circles). The direct experimental input is sharper than the model for the heaviest and sub-dominant components ($^8$B and $^7$Be), which burn closer to the Sun's core. In contrast, the dominant and lighter components (pp and pep) remain primarily constrained by the SSM. The second abyss corresponds to today's ~20× gap between experimental precision and predictions for the most important pp component (>90%), thus underscoring the need for new solar neutrino detection. This increases to ~40× if compared to the current dark matter experiments' results. Although less prominent in our Sun, the most common fusion in the Universe is the CNO, whose direct exploration with neutrinos is only just at the beginning.

This powerful idea drove the LENS project **[28]**, which operated in an unforgiving radiogenic background, despite the difficulty of loading enough indium without extinguishing the light. The self-radioactivity of indium is severe: natural indium's own β-decay is the dominant — directly comparable with the $^{14}$C wall — and largely irreducible with today's solar neutrino detection paradigm. In that battle, scintillator loading was pushed to record levels at precisely the regime where LiquidO's opaque scintillation is designed to excel, because heavy loading sharpens the topology and per-event position that could, in principle, lift the indium signal out of its own radioactivity. This is a design goal, and whether LiquidO can deliver the required rejection is exactly a question to be addressed by the ongoing demonstration programme. While nobody has ever seen solar neutrinos with indium, LiquidO is arguably today the detection framework to attempt indium at the scale solar detection needs — the very context of six-decade-old detector technology has not fully succeeded. Demonstrating this potential is the next experimental step.

## The multi-messenger ecosystem — the astrophysical state of the art

Since neutrinos have so far been the Sun's "shy" messenger — due to propagation uncertainties and detection capabilities — we shall begin with the eloquent ones: acoustic waves and light. The solar interior is best read zone by zone (**Figure 2**): the convective envelope (≳0.71 R⊙), the radiative interior (from the centre to ~0.71 R⊙) and, nested within it, the nuclear-burning region or core, whose extent depends on the reaction — heavier elements require higher depth in the core to burn.

Escaping photons last interact at the photosphere; acoustic modes traverse all radii, including the core, with degrading resolution in the innermost regions, exactly where neutrinos are produced. Among the acoustic oscillations (**Figure 2**), the most spatially informative are the acoustic local-modes diagnostics (three-dimensional, but mainly in the near-surface and upper convection zone), distinct from the global-modes (whose frequencies constrain the spherically averaged radial structure at all radii, and whose rotational splittings constrain the axisymmetric rotation profile $\Omega(r,\theta)$) and the g-modes, or so-called "gravity" modes, are predicted by standard theory but remain unconfirmed **[10,29]**. No other established messenger provides comparable direct access to the solar interior.

Helioseismology, the study of the Sun's oscillations by ground networks (GONG, BiSON) and space missions (SOHO, SDO), has inferred i) the radial sound-speed profile to a few permille across much of the interior, and ii) the convection-zone depth and envelope helium abundance with a precision no stellar model had matched **[10,20]**. Space observatories now scrutinise the surface, atmosphere and wind in remarkable detail, from SOHO and SDO to Parker Solar Probe and Solar Orbiter **[30]**. Its very success created one of the field's sharpest problems. The sequence was the reverse: seismology first established precise structural constraints; older, higher-abundance SSMs matched them well; then the downward revision of the photospheric C, N, O abundances in the 2000s produced models that disagree more strongly with those pre-existing constraints **[8,31]**. This is often summarised as a high/low metallicity dispute. Still, the picture remains genuinely unsettled, as some analyses raise the metallicity and argue that agreement is restored **[32]**. In contrast, others find that higher abundances alone do not resolve the coupled opacity, mixing, and transport problems **[33]**. The cleanest neutrino handle is high-precision CNO direct detection — one of the most direct, independent constraints on the core C+N abundance — though model-dependent **[13,34]**. While the outside-in picture is exquisite over most of the interior, the piece it cannot supply (the direct nuclear source terms of the core) is precisely for the neutrinos to write.

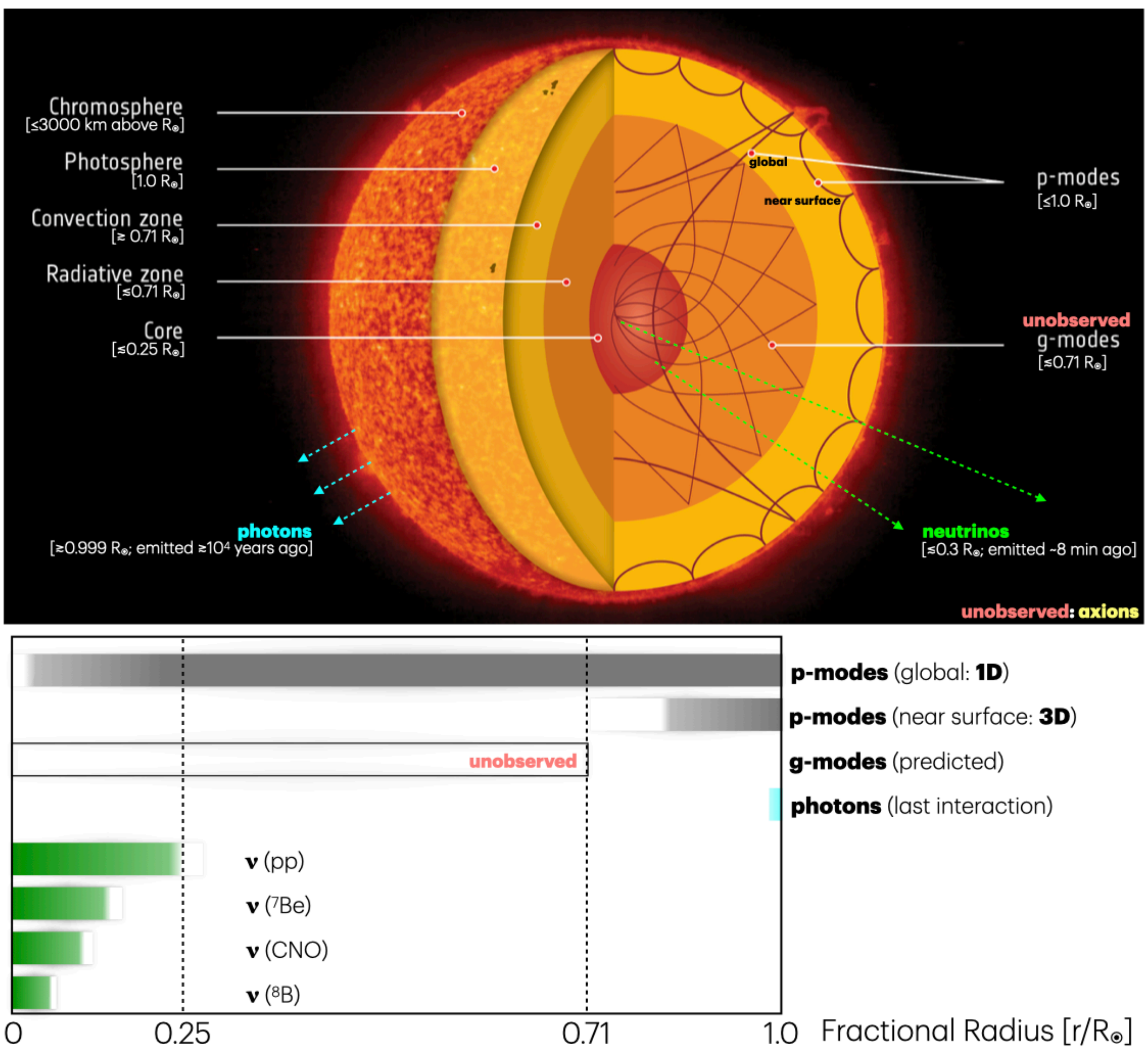

**Figure 2 | The multi-messenger ecosystem of the solar interior.** The 3D cross-section of the Sun **(top)** illustrated the regions below the *photosphere* (1.0 R⊙): the *convective* envelope (above 0.71 R⊙), the *radiative* interior (below 0.71 R⊙) and, nested within it, the graded nuclear-fusion-burning *core* (dominant below ~0.25 R⊙). The extent of the core depends on the fusion reaction. *Inside-out messengers* originate from the core: ***neutrinos*** (travel to Earth in ~8 minutes) and, hypothetically, ***axions*** (still unobserved). Neutrino production is graded and component-dependent: pp/pep are broadly distributed across the core (≤0.25 R⊙), while heavier $^{7}$Be, CNO, and $^{8}$B are progressively more centralised (≤0.15 R⊙). *Outside-in probes*, detected at or near the surface, are the escaping ***photons*** (last interaction at the photosphere, ≥0.999 R⊙; emitted from deeper layers, up to $10^5$ years earlier) and ***acoustic oscillations***. Global ***p-modes*** probe the spherically averaged structure at all radii and rotation $\Omega(r,\theta)$, while local diagnostics (shallow acoustic paths) map the near-surface/upper convection zone in 3D. The ***g-modes*** (predicted buoyancy waves, unobserved) are trapped in the radiative interior and core, evanescent in the convective envelope, and thus extraordinarily faint — as predicted. Each carrier reads the Sun differently **(bottom)**: the production and probing kernels for each messenger are illustrated, with neutrinos mainly sensitive to the Sun's core. This implies that a high degree of complementarity and synergy is expected when all messengers can be observed simultaneously. [Sun's image is a modified version based on the SOHO image. Credit: SOHO (ESA & NASA)].

More subtle is the physical decomposition of the radiative interior. Its mechanical structure (sound speed, density, and the convection-zone base) is tightly constrained seismically; what remains non-unique is the decomposition into temperature, mean molecular weight, composition, opacity and transport **[8,10,20]**. Tellingly, the model-to-seismology sound-speed discrepancy peaks near the base of the convection zone **[8,31]**. The messenger born to read this region, the internal gravity mode, has its amplitude concentrated in the radiative interior and core but is evanescent in the convectively unstable envelope and thus extraordinarily faint; a claimed signature remains unconfirmed after nearly two decades, alongside later claims and their critiques **[35]**. The synergy of *multi-messenger heliotomography*[2] is strongest here: a joint analysis — a forward-modelled, regularised inference in which composition, opacity, diffusion, mixing and nuclear rates are varied and compared simultaneously with seismic and neutrino data — may break today's degeneracies that neither messenger resolves alone **[8,19,20,34]**. Moreover, the inferential framework for such a programme already exists: Bayesian reconstruction using helioseismic and neutrino data has been demonstrated **[17,18]**, though it has been limited by the experimental precision available at the time and by the computational cost of generating SSM realisations. Today's ongoing computing revolution is poised to lead to major improvements. Indeed, providing light to this darkness is a quantitative, multi-messenger goal — one whose methodological foundations are laid and whose rate-limiting step is soon the experimental input.

## Neutrino detection: state of the art

Today's solar neutrino *state-of-the-art*, led by transparent detectors, can be classified into two families. First, Borexino **[11]** (liquid scintillator) has measured all major solar-neutrino families (except hep), including observations of both the pp chain and the CNO signal, with component-dependent precision (**Figure 1**). And second, experiments sensitive only (or mainly) to the higher-energy $^{8}$B flux. The latter is typically accessible to essentially all solar-capable detectors. Our knowledge is dominated by nearly three decades of Super-Kamiokande data, whose few percent precision contrasts with a much poorer SSM prediction (~15%). For neutrinos to become a powerful astrophysical tool, they must reach or surpass SSM precision across the entire spectrum — both pp and CNO — which today requires more than one order-of-magnitude improvement, especially at the lowest-energy end (pp).

As byproducts of their main scientific goals, JUNO **[23]** and other large future projects, including Hyper-Kamiokande and DUNE, will mainly target $^{8}$B **[19,36]**, with background permitting slightly beyond. In addition, two concepts follow Borexino's ultra-radiopure organic scintillators: THEIA and Jinping; thus, their ultimate expected floor is set by $^{14}$C, thus limiting pp neutrino detection. Most recently, the largest dark-matter experiments (both xenon-based), XENONnT **[37]** and PandaX **[38]**, have consistently demonstrated their ability to probe the pp region as well. While their current precision is not competitive, the challenges ahead include scaling to the next generation, systematics and backgrounds **[39]**, and $^{14}$C contamination, which remains a delicate issue. All of those cases are impeccably transparent and pure detector systems, typically requiring deep underground sites with overburden depths (between ~0.7 to ~2.0 km) to suppress cosmic-muon-induced cosmogenic backgrounds.

### The next solar-astrophysics neutrino flagship experiment?

Anticipating today's inflexion point in the potential of solar neutrino astrophysics, our team has been developing, for almost a decade, a new detector to address this challenge. Building on the LENS heritage, the concept relies on an indium-doped target with double-flash detection to beat $^{14}$C once empowered by LiquidO's topological imaging and spatial resolution. Cosmogenic backgrounds permitting, this may enable sites near the surface to target permille statistical precision, particularly for the pp component, with solar-neutrino astrophysics as the primary scientific driver. This flagship experiment candidate is still under design, though. Of course, this same precision is not directly applicable to all derived astrophysical observables.

[2] Multi-messenger heliotomography refers to the ability to combine all messenger data to reconstruct 3D models of the Sun's structure and physical features, including its stability.

The SuperChooz project (based in Europe, France), under exploration since 2017–2019, was first presented in full at CERN in 2022 **[40]** and is the first large-scale embodiment of opaque-scintillator imaging: a large LiquidO detector (~10 kton) reusing the historic galleries of the EDF Chooz-A reactor **[41]** — one of the first reactors in Europe — now under dismantling. Indeed, the location is not by chance, as its founding mission was reactor physics via inverse beta decay, whose background is already well known in situ by the Double Chooz experiment **[42]**.

However, additionally doped with indium (up to 10% by mass under consideration and optimisation), along with its event-by-event imaging, could enable simultaneous detection of both reactor antineutrinos and solar neutrinos — one detector, two missions — directly bridging the controlled source and the wild one, including a unique direct cross-check of JUNO measurements. With sufficient statistics, SuperChooz could also use the Sun to probe fundamental particle physics that may manifest as smaller corrections to the overall neutrino propagation model. Here, precision in the few-MeV survival-probability transition tests the conventional matter effect **[43]** — a direct probe of the Sun's density profile and possible new interactions — thus opening a discovery-sensitive window. A spatial map of the electron-density profile is a key input to the resulting heliotomography that combines all elements.

SuperChooz could aim to measure, in real time, the pp chain and the CNO cycle, including most of their sub-components, via full spectral extraction, thereby obtaining a thermometric reading of the solar core and confronting the abundance problem through CNO spectroscopy. The goal is to reach and, where possible, surpass today's SSM flux uncertainties for both the pp chain and the CNO cycle **[9]**. As each flux probes a different depth and rung of the fusion ladder, one can sharpen the neutrino-luminosity test by comparing the Sun's present nuclear power (neutrinos in real time) with its delayed photon luminosity (reflecting energy produced up to $10^5$ years ago): a unique solar-luminosity test **[44]** of the quasi-stationary thermodynamic stability of the Sun across radically different transport timescales, with the potential to provide unique constraints on nuclear energy generation and additional freely escaping energy-loss channels such as axion emission **[45]**. The final ambition is to constrain the relative spatial distributions of the different branches of hydrogen burning within a model-dependent framework. The sense in which heliotomography is meant here: aspirational, model-dependent, and defined by production kernels and the survival probability rather than a direct image.

## Proving the indium dream in a reactor-based demonstrator

Before SuperChooz can watch the sky, the essential step is to demonstrate its feasibility, including key challenges ahead. The most important limitation is that there is no readily available terrestrial MeV-neutrino source capable of reproducing the solar-neutrino spectrum. That is why the most ambitious solar experiments in history, such as SNO, ultimately had to step into the unknown **[3]**: SNO's decisive deuterium measurement mainly relied on an a priori estimate, as the preceding experiment, run earlier by *Reines* et al. using a reactor in the 1970s, had proven inconclusive **[46]**. The only known alternative so far is the use of intense monoenergetic neutrino sources — typically reactor-activated, such as the $^{51}$Cr source used in the GALLEX, SAGE, and BEST experiments, to probe selected channels and validate detector response **[47]**.

Demonstrating SuperChooz's feasibility is the original birth lineage of the CLOUD experiment **[48]**, thus designed to explore the indium–LiquidO scenarios upon completion of the AntiMatterOTech innovation programme, funded by the EIC and UKRI **[49]**. There, the Chooz reactor serves as a "test-beam" facility to demonstrate detection, notably including the observation of subdominant reactor neutrinos beyond the dominant fission byproducts. Indeed, neutrinos can be emitted (at a much smaller flux) from reactor-activated materials, and chromium is abundant **[47]**. No such activation-neutrino has ever been detected. However, almost sixty years ago, almost simultaneously with the first observation of reactor antineutrinos by *Reines* et al., *Davis* exposed his neutrino-only-sensitive detector to a reactor and obtained a critical “null result” **[50]**. This key observation established the $\nu/\bar{\nu}$ distinction. So, CLOUD will revisit this for the first time since using indium detection.

One of CLOUD's main goals is to probe and challenge the LENS background model and explore all possible background scenarios that have not been considered. Hence, the setting is deliberately harsh: a small detector, close to a reactor, almost at the surface — each condition is enough to sink most transparent detectors today. If LENS' background model was correct, CLOUD could measure enough pp-chain neutrinos for a $\geq 5\sigma$ observation over a year's exposure against a small (a priori unknown) activation-neutrino background from the reactor **[48]**. Either observation would be a first of its kind.

## From heliotomography to geotomography

The “holy” inverse-beta-decay interaction also has limitations, such as its energy threshold (1.8 MeV), defined by the mass difference between the proton and the neutron. This is an issue for geoneutrinos, i.e., antineutrinos from natural radioactivity within the Earth — a channel pioneered by KamLAND in 2005 **[51]**, with important contributions now from Borexino, SNO+ and JUNO. In a LiquidO detector, the presence of antimatter has an extra tagging advantage: the unambiguous positron annihilation signature. This detection principle is expected to significantly improve reactor antineutrino detection by reducing backgrounds, a key design criterion for the CLOUD experimental setup. However, positron ID is not enough.

Pioneered by our teams, a new interaction may be necessary, benefiting from the positron ID to lower the threshold to ~1.2 MeV and thereby enable geoneutrino detection. If successful, this opens a route to unobserved Earth's “missing potassium” **[52]**, for which copper is, as far as we know, the only practical target — again, a challenge only a LiquidO-type detector may attempt. Of course, so close to a reactor, SuperChooz cannot detect geoneutrinos on its own. Still, CLOUD may use the reactor core to mimic Earth's geoneutrinos for a first observation — a critical demonstration step. CLOUD is thus poised to read a reactor core as a proxy for the Earth: an imaging principle that turns the opaque interiors of the reactor, Earth, and Sun into a legible object, extending its reach — one day, maybe — toward a broader *geotomography*.

## After 60 years, here comes the Sun

The pioneers, *Davis et al.*, located a tank of dry-cleaning fluid in a mine in South Dakota to look inside our star. A move many would have called madness led to new physics — fixing neutrino propagation about sixty years later — and to a multidisciplinary field that shaped generations of scientists. Much has changed since. Indeed, many neutrino physicists now wonder whether solar neutrinos are effectively finished, and many astronomers, tired of waiting, have set aside the neutrino's “*el Dorado*”, which had so far mostly corroborated conclusions the SSM reached with modest neutrino input. Yet after this long, worthwhile latency — encompassing many groundbreaking discoveries — we argue the era of solar neutrino astrophysics may now restart, upon defeating two abysses so far.

Sixty years on, merely sixty days of reactor antineutrinos — designed with decades of worldwide wisdom — have repaid the first debt: the neutrino's journey from the Sun will soon be charted to the precision the SSM demands. That closes the first abyss. The second, associated with the fact that no instrument yet reads solar neutrinos precisely enough to exploit that journey, remains open; a concrete instrumentation challenge whose solution — one at least — may be in the making if demonstrated.

This exploration makes the intimate interplay between solar and reactor MeV neutrino detection matter more important than ever: the wildness of the Sun and the mysteries of the Earth may be read only thanks to the unprecedented detection precision that reactors can experimentally probe in situ. This synergy is also poised to extend the multi-messenger revolution to other astrophysical objects, for example, via supernova neutrino spectroscopy, relying on the same detection principle. The outcome of this operation is as simple as powerful: the Sun stops being an instrument and becomes our object. So, the neutrino — the long-promised new messenger — can join sound to complete a multi-messenger heliotomography of our star. After so long, the next sixty years can belong to the Sun, if — and only if — the field builds the instruments up to the task.

## Acknowledgements

The authors would like to express their most sincere gratitude to *M. C. González-García* for her deep insight and important suggestions, which have helped improve and contextualise the manuscript. AC, MC, JH, DN, and FY acknowledge the core contribution of the EIC Pathfinder project "AntiMatter-OTech" (ID: 101047028), funded by the EIC and UKRI, which provides the demonstration basis for the LiquidO-based experimental context for exploring future projects, such as CLOUD and SuperChooz. AC and FY acknowledge support from the CNRS, Université Paris Saclay and Nantes Université, France. SD acknowledges support from the Istituto Nazionale di Fisica Nucleare (INFN), Italy. HN acknowledges financial support provided by Brazilian Funding Agencies CNPq and CAPES. JMC acknowledges support from the MICINN through the grant "DarkMaps" PID2022-142142NB-I00 and from the European Union through the grant "UNDARK" of the Widening Participation and Spreading Excellence Programme (project number 101159929). The use of AI has been restricted mainly to editorial support and validation, including unbiased selection of relevant references across the field's literature.